\documentclass[english,showpacs,preprint,superscriptaddress]{revtex4-1}

 \UseRawInputEncoding  
 
\usepackage{babel}
\usepackage[T1]{fontenc}      
\usepackage{mathrsfs}         
\usepackage{bm}
\usepackage{color, xcolor}   
 
\usepackage{lmodern}     

\usepackage{amsmath,amsfonts,amssymb}
\usepackage{esint}     
\usepackage{extarrows}

\usepackage[ruled,linesnumbered]{algorithm2e}
\usepackage{algpseudocode}

\usepackage{graphicx}   
\usepackage{float}
\usepackage{tikz}       
\usepackage{palatino}
\usetikzlibrary{arrows,shapes,chains}

\usepackage{subfig}     
\usepackage{booktabs}  
\graphicspath{{papers/All/figure/}}

\usepackage[justification=centering]{caption}
\usepackage{siunitx}

\usepackage{indentfirst}
\usepackage{setspace}

\makeatletter

\usepackage[title]{appendix}

\usepackage{hyperref}  
\hypersetup{
    breaklinks = true,
    colorlinks = true,
    citecolor = {blue},
    urlcolor = {blue},
    linkcolor = {blue}
} 

\usepackage{cleveref}

 \global\long\def\calK{\mathcal{K}}
 
 \global\long\def\calM{\mathcal{M}}

 \global\long\def\calMh{\hat{\mathcal{M}}}

 \global\long\def\calRh{\hat{\mathcal{R}}}

 \global\long\def\scrf{\mathscr{f}}

 \global\long\def\scrfh{\hat{\mathscr{f}}}

 \global\long\def\frakC{\mathfrak{C}}

 \global\long\def\frakCh{\hat{\mathfrak{C}}}

 \global\long\def\frakm{\mathfrak{m}}
 \global\long\def\frakn{\mathfrak{n}}

 \global\long\def\rmP{\mathrm{P}}
 
 \global\long\def\rmR{\mathrm{R}}
 
 \global\long\def\rmT{\mathrm{T}}

 \global\long\def\rmY{\mathrm{Y}}

 \global\long\def\rmd{\mathrm{d}}
 \global\long\def\rme{\mathrm{e}}

 \global\long\def\rmv{v}

 \global\long\def\rmvh{\hat{v}}

 \global\long\def\bsB{\boldsymbol{B}}
 \global\long\def\bsC{\boldsymbol{C}}
 
 \global\long\def\bsE{\boldsymbol{E}}

 \global\long\def\bsr{\boldsymbol{r}}

 \global\long\def\bsv{\boldsymbol{v}}

 \global\long\def\bbN{\mathbb{N}}

 \global\long\def\nh{\hat{n}}
 \global\long\def\uh{\hat{u}}

 \global\long\def\vb{\bsv_b}
 \global\long\def\v{\bsv}

 \global\long\def\vvbth{\rmvh_{ab}} 

 \global\long\def\r{\bsr}
 \global\long\def\rmvhb{\rmvh_b}

 \global\long\def\Ylm{\rmY_l^m}

 \global\long\def\Pl{\rmP_l}

 \global\long\def\ddt{\frac{\partial}{\partial t}}

 \global\long\def\ddbfv{\nabla_{\v}}

 \global\long\def\erf{\mathrm{erf}}
 \global\long\def\rmGamma{{\Gamma}}
 \global\long\def\e{\rme}
 \global\long\def\IIFI{\mathrm{_2 F_1}}
 \global\long\def\IFI{\mathrm{_1 F_1}}
 
 \global\long\def\navth{\frac{n_a}{\vath^3}}
 \global\long\def\nbvth{\frac{n_b}{\vbth^3}}

 \global\long\def\fl{f_l}
 \global\long\def\fo{f_0}

\global\long\def\fh{\hat{f}}

\global\long\def\Fh{\hat{F}}

 \global\long\def\fho{\fh_0}

 \global\long\def\dfho{\frac{\mathrm{\partial} \fho}{\mathrm{\partial} \hat{\rmv}}}
 \global\long\def\ddfho{\frac{\mathrm{\partial}^2 \fho}{\mathrm{\partial} {\hat{\rmv}}^2}}

\global\long\def\Fh{\hat{F}}
\global\long\def\Hh{\hat{H}}
\global\long\def\Gh{\hat{G}}

 \global\long\def\Fho{\Fh{_0}}
 \global\long\def\Hho{\Hh{_0}}
 \global\long\def\Gho{\Gh{_0}}

 \global\long\def\dHho{\frac{\partial \Hho}{\partial \vvbth}}

 \global\long\def\dGho{\frac{\partial \Gho}{\partial \vvbth}}
 \global\long\def\ddGho{\frac{\partial^2 \Gho}{\partial {\vvbth}^2}}

  \global\long\def\calMjo{{\calM}{_{j,0}}}
  \global\long\def\calMoo{{\calM}{_{0,0}}}

  \global\long\def\calMhjo{{\calMh}{_{j,0}}}

  \global\long\def\calRhjo{{\calRh}{_{j,0}}}
  
  \global\long\def\calRhab{{\calRh}{_{ab}}}

  \global\long\def\calRhabjo{{\calRhab}{_{j,0}}}
  \global\long\def\calRhaboo{{\calRhab}{_{0,0}}}

\global\long\def\calRhabrs{{\calRh}{_{ab}^{r,s}}}

\global\long\def\calRhabrsjo{\calRhabrs_{j,0}}

\global\long\def\IhSabrs{{\hat{I}_S}{_{ab}^{r,s}}}

\global\long\def\IhSabrsjo{\IhSabrs{_{j,0}}}

\global\long\def\IhCabrs{{\hat{I}_C}{_{ab}^{r,s}}}

\global\long\def\IhCabrsjo{\IhCabrs{_{j,0}}}

\global\long\def\IhDabrs{{\hat{I}_D}{_{ab}^{r,s}}}

\global\long\def\IhDabrsjo{\IhDabrs{_{j,0,0}}}
\global\long\def\IhDabrsjo{\IhDabrs{_{j,0}}}

 \global\long\def\Ko{\calK_0}

 \global\long\def\scrfI{\scrf{_1}}

 \global\long\def\scrfI0{\scrf{_1^0}}

 \global\long\def\scrfhI{\scrfh{_1}}

 \global\long\def\scrfhI0{\scrfh{_1^0}}

 \global\long\def\Bn{\bsB}
 \global\long\def\Cn{\bsC}              
 \global\long\def\En{\bsE}

 \global\long\def\IiFho{I_{i,0}}

 \global\long\def\JiFho{J_{i,0}}

 \global\long\def\lnAab{\ln{ \Lambda_{ab}}}

 \global\long\def\Gab{\Gamma_{ab}}

 \global\long\def\cola{\frakC}

 \global\long\def\colla{{\frakC{_l}}}

 \global\long\def\colhla{{\frakCh{_l}}}
 \global\long\def\colhoa{{\frakCh{_0}}}

 \global\long\def\colab{\frakC_{ab}}

 \global\long\def\colhlab{{\frakCh{_l}}_{ab}}
 \global\long\def\colhoab{{\frakCh{_0}}_{ab}}

 \global\long\def\CHh{C_{\Hh}}
 \global\long\def\CGh{C_{\Gh}}

\global\long\def\uzar{u_{a_r}}

\global\long\def\uzhar{\uh_{a_r}}

\global\long\def\uzhbs{\uh_{b_s}}

 \global\long\def\vath{\rmv_{ath}}
 \global\long\def\vabth{\rmv_{abth}}

 \global\long\def\vathr{\rmv_{{ath}_r}}

 \global\long\def\vhathr{\rmvh_{{ath}_r}}

 \global\long\def\vbth{\rmv_{bth}}

 \global\long\def\vhbths{\rmvh_{{bth}_s}}

\global\long\def\nhbs{\nh_{b_s}}

\global\long\def\nar{n_{a_r}}
\global\long\def\nhar{\nh_{a_r}}

 \global\long\def\nuTab{\nu_T^{ab}}
 
 \global\long\def\nuTa{\nu_T^{a}}

\global\long\def\NKa{N_{K_a}}
\global\long\def\NKb{N_{K_b}}

\global\long\def\vvbth{\rmvh_{ab}}

 \global\long\def\FIG#1{~\ref{#1}}
 \global\long\def\EQ#1{~(\ref{#1})}
 \global\long\def\EQo#1{(\ref{#1})}

 \global\long\def\SEC#1{~\ref{#1}}
 \global\long\def\APP#1{~\ref{#1}}

\makeatother

\begin{document}
 
\title{General relaxation model for a homogeneous plasma with spherically symmetric velocity space}

\author{Yanpeng Wang}
\email{Corresponding author. E-mail: tangwang@mail.ustc.edu.cn}
\affiliation{School of Nuclear Sciences and Technology, University of Science and Technology of China, Hefei, 230026, China}

\author{Shichao Wu}
\email{Co-first author. E-mail:wusc@jou.edu.cn}
\affiliation{School of Science, Jiangsu Ocean University, Lianyungang, 222005, China}




\author{Peifeng Fang}
\affiliation{School of Physics and Optoelectronic Engineering, Anhui University, Hefei 230601, China}



\begin{abstract}
  A kinetic moment-closed model (KMCM), derived from the Vlasov-Fokker-Planck (VFP) equation with spherically symmetric velocity space, is introduced as a general relaxation model for homogeneous plasmas. The closed form of this model is presented by introducing a set new functions called $R$ function and $R$ integration. This nonlinear model, based on the finitely distinguishable independent features (FDIF) hypothesis, enables the capture of the nature of the equilibrium state. From this relaxation model, a general temperature relaxation model is derived when velocity space exhibits spherical symmetry, and the general characteristic frequency of temperature relaxation is presented.
  
  \
   
   
\end{abstract}

\maketitle

 \UseRawInputEncoding




 
\begin{spacing}{0.35}  

\section{Introduction}
\label{Introduction}     

The evolution of fusion plasmas over time can be effectively depicted by the Vlasov-Fokker-Planck\cite{Rosenbluth1957,  Vlasov1968} (VFP) equation in combination with the Maxwell equations. Nevertheless, except for a few specific cases such as thermodynamic equilibrium\cite{wang2024Relaxationmodel, Boltzmann1966}, the VFP equation typically exhibits significant nonlinearity\cite{Mintzer1965}. Solving the VFP equation, either analytically or numerically, often requires certain assumptions to simplify the equation\cite{wang2025TransportI, Schunk1977}. Furthermore, numerical solutions to the VFP equation are generally classified into two approaches: direct discretization methods\cite{Thomas2012, wang2024Aconservative} and moment methods\cite{Grad1949, wang2025TransportI}. 

The moment methods typically transform the VFP equation into a set of nonlinear equations, which often lack closure\cite{Mintzer1965, wang2025TransportI} and demand some assumptions to be enclosed. 
Traditionally, the near-equilibrium assumption is the widely adopted and effective approach in plasmas physics, such as the Grad's moment theory\cite{Grad1949} derived from VFP equation based on Hermite polynomial expansion (HPE), as well as the traditional low-order moment theories which are based on the Chapman-Enskog expansion\cite{Chapman1916The, Chapman1953, Burnett1935} (essentially a Taylor expansion), including the magnetohydrodynamic equations, two-fluid equations, and so forth\cite{Freidberg2014}. However, the near-equilibrium assumption enforce the expansions of the distribution function into an orthogonal series around a local Maxwellian, also leading to inadequate convergence in highly non-Maxwellian system\cite{Schunk1977}, including both the Grad and Chapman-Enskog approaches.

In our previous works\cite{wang2024Relaxationmodel, wang2024Aconservative}, another assumption, namely finitely distinguishable independent features (FDIF) hypothesis, has been introduced to enclose the moment equations\cite{wang2025TransportI}. Based on FDIF hypothesis, a novel framework\cite{wang2024higherorder} is provided for addressing the nonlinear simulation in fusion plasmas. In this framework, spherical harmonics expansion (SHE) is utilized for the angular coordinate in velocity space, and the King function expansion (KFE) method is employed for the speed coordinate, which has been demonstrated to be a moment convergent method\cite{wang2024Aconservative}. Reference\cite{wang2024Relaxationmodel} has presented a relaxation model for homogeneous plasmas in scenario with shell-less spherically symmetric velocity space. In this study, we will offer a further development of the relaxation model for general spherically symmetric velocity space, which may incorporate shell structures\cite{Min2015, wang2024Relaxationmodel}. Based on this general nonlinear relaxation model, a general temperature relaxation model will also be presented when velocity space exhibits spherical symmetry.

The following sections of this paper are organized as follows. Sec.\ref{Theoretical formulation} provides an introduction to the VFP equation and its spectrum form. Sec.\SEC{General relaxation model for homogeneous plasmas} deliberates on the general relaxation model in cases where velocity space exhibits spherical symmetry, including the closure relations. Then, a general temperature relaxation model is derived from the provided relaxation model in  Sec.\SEC{Temperature relaxation model}. Finally, a summary of our work is presented in Sec.\SEC{Conclusion}.

\end{spacing}

\begin{spacing}{0.7}   

\section{Theoretical formulation}
\label{Theoretical formulation}

\subsection{Vlasov-Fokker-Planck equation}
\label{Vlasov-Fokker-Planck equation}

The relaxation of Coulomb collision in fusion plasmas can be described by the VFP\cite{Rosenbluth1957,  Vlasov1968} equation for the velocity distribution functions, denoted as $f=f \left(\r, \v,t \right)$ for species $a$, in physics space $\r$ and velocity space $\v$, reads
  \begin{eqnarray}
      \ddt f + \v \cdot \nabla {f} + \frac{q_a}{m_a} \left(\En + \v \times \Bn \right) \cdot \ddbfv f  &=& \cola ~.\label{VFP}
  \end{eqnarray} 
Here, symbols $m_a$ and $q_a$ respectively denote the mass and charge of species $a$. Vectors $\En$ and $\Bn$ respectively represent the electric field intensity and magnetic flux density. Function $\cola$ is the Coulomb collision operator of species $a$, which encompasses both its self-collision effect and the mutual collision effect between species $a$ and background species, as described by
  \begin{eqnarray}
      \cola \left(\r, \v,t \right) &=& \sum{_{b=1}^{N_s}} \colab  ~,\label{cola}
  \end{eqnarray}
where $N_s$ denotes the total number of plasmas species and $\colab$ represents the mutual FPRS collision operator\cite{wang2024Relaxationmodel,Shkarofsky1963,Shkarofsky1967} between species $a$ and species $b$, given by
  \begin{eqnarray}
      \colab \left(\r, \v,t \right) &=& \Gab 
      \left [4 \pi m_M F f + \left(1-m_M \right) 
      \ddbfv H \cdot \ddbfv f + \frac{1}{2} \ddbfv \ddbfv G : \ddbfv \ddbfv f
      \right]~.  \label{FPRS}
  \end{eqnarray}
In this content, $ \Gab=4\pi \left(\frac{q_a q_b}{4\pi \varepsilon_0 m_a} \right)^2 \lnAab$, the mass ratio $m_M=m_a/m_b$. Here, symbols $m_b$ and $q_b$ respectively represent the mass and charge of species $b$. Parameters $\varepsilon_0$ and $\lnAab$ correspond to the dielectric constant of vacuum and the Coulomb logarithm\cite{Huba2011}.
Function $F=F(\vb,t)$, denoting the background distribution function of species $b$. The Rosenbluth potentials are integral functions of $F$, which can be expressed as follows
  \begin{eqnarray} 
    H(\r, \v,t) &=& \int \frac{1}{\left|\v-\vb \right|} F\left(\r, \vb,t \right) \rmd \v _b, \label{H}
    \\
    G(\r, \v,t) &=& \int \left|\v-\vb \right| F\left(\r, \vb,t \right) \rmd \v _b ~. \label{G}
  \end{eqnarray} 

This paper exclusively focuses on the general scenario where the velocity space exhibits spherical symmetry in a spatially homogeneous plasma. It can be proved\cite{wang2025TransportI} that the spatial convection terms (second term), electric and magnetic field effect terms (third term) in Eq.\EQ{VFP} will vanish exactly when velocity space exhibits spherical symmetry. Therefore, the VFP equation reduces to
  \begin{eqnarray}
      \ddt f \left({\mathbf{v}},t \right)  &=& \cola ~.\label{VFP0D1V}
  \end{eqnarray} 
Hereafter, we disregard the symbol $\r$ in this homogeneous plasmas scenario.

\subsection{VFP spectrum equation}
\label{VFP spectrum equation}

By employing spherical harmonics expansion \cite{Johnston1960,Bell2006} (SHE) within a spherical-polar coordinate ${\mathbf{v}}(v, \theta,\phi)$, the distribution function of species $a$ can be expended as
  \begin{eqnarray}
      f \left({\mathbf{v}},t \right) &=& {\sum_{l=0}^{\infty} } {\sum_{m=-l}^l } {f{_{l}^m}} \left(v ,t \right) {\mathrm{Y}_l^m} \left(\mu, \phi \right) , \label{fhsph}
  \end{eqnarray}
where $v = |\v|$, $\mu=\cos{\theta}$ and ${\mathrm{Y}_l^m}$ represents the spherical harmonic\cite{Arfken1971} without the normalization coefficient, $N_l^m=\sqrt{\frac{2l+1}{4\pi} \frac{(l-m)!}{(l+m)!}}$.  Then, we can obtain the VFP spectrum equation\cite{wang2024Relaxationmodel} in spherically symmetric velocity space, derived from Eq.\EQ{VFP0D1V}. For order of $l$, that is
  \begin{eqnarray}
      \ddt \fl \left(\rmvh,t \right) &=& \delta_l^0 \colla \left(\rmvh,t \right)  , \label{VFPl}
  \end{eqnarray}
where $\delta_l^0$ denotes the Kronecker symbol. Eq.\EQ{VFPl} indicates that harmonics with order $l \ge 1$ or $m \ne 0$ will vanish due to the spherical symmetry of velocity space\cite{Tzoufras2011}. For convenience, we adopt the normalized speed as the dependent variable, for species $a$ denoted as $\rmvh = v / \vath$. Moreover, when velocity space exhibits spherical symmetry, $\Ylm \to \Pl$. In this paper, we omit the superscript $m$ because its value is always zero. For instance, we utilize ${f_l}$ instead of ${f_l^0}$. 
The amplitude $\fl$ is zero when $l \ge 1$ and $\fo$ is non-negative, which can be computed through the inverse transformation of SHE, reads
   \begin{eqnarray}
       \fl \left(\rmvh, t \right) &=& \delta_l^0 \frac{1}{(N_l)^2} \int_{-1}^1 \int_0^{2 \pi} f \left (\rmv / \vath, \mu, \phi, t \right)   \Pl (\mu) \mathrm{d} \phi \rmd \mu, \label{flf}
   \end{eqnarray}

The $l^{th}$-order amplitude of FPRS collision operator\cite{wang2024Relaxationmodel} of species $a$ in Eq.\EQ{VFPl} can be expressed as
  \begin{eqnarray}
      \colla \left(\rmvh,t \right) &=& \delta_l^0 \navth \colhla \left(\rmvh,t \right). \label{Chla}
  \end{eqnarray} 
The parameter $n_a = \int_{\v} f \rmd \v$, and  $\vath$ respectively denote the number density and thermal velocity of species $a$, which will be further elaborated in Sec.\SEC{Transport equation}. 
The $l^{th}$-order normalized FPRS collision operator of species $a$ is defined as
  \begin{eqnarray}
      \colhla \left(\rmvh,t \right) &=& \delta_l^0 \sum{_{b=1}^{N_s}} \nbvth \Gab \colhlab. \label{Cla}
  \end{eqnarray} 
Similarly, the parameter $n_b$ and $\vbth$ denote the number density and thermal velocity of species $b$. The $l^{th}$-order normalized amplitude of FPRS collision operator for species $a$ colliding with species $b$ can be represented as\cite{wang2024Relaxationmodel}
  \begin{eqnarray}
      \colhlab \left(\rmvh,t \right) = \delta_l^0 4 \pi \Gab \left [m_M \Fho\left(\vvbth,t \right) \fho + \CHh \dHho 
      \dfho + \frac{2 \CGh}{\vvbth \rmvh}  \dGho \dfho + 
      \CGh \ddGho  \ddfho \right] 
      , \label{collaba}
  \end{eqnarray}
where $\vvbth = \vabth \rmvh$, the thermal velocity ratio $\vabth = \vath/\vbth$ and the coefficients,
  \begin{eqnarray}
      \CHh= \frac{1-m_M}{\vabth} , \quad 
      \CGh= \frac{1}{2 \left(\vabth \right)^2}
      ~. \label{CFHGh}          
  \end{eqnarray}
  
In above equation\EQ{collaba}, the normalized amplitude of distribution function can be calculated as follow
  \begin{eqnarray}
      \fho \left(\rmvh,t \right) &=& \vath^3 n_a^{-1} \fo \left(\rmvh,t \right)~. \label{fh0}
  \end{eqnarray}
The normalized amplitudes of the background distribution function and its relative Rosenbluth potentials can be formulated as follows
\begin{eqnarray}
    \Fho \left(\vvbth,t \right) &=&  \vbth^3 n_b^{-1} F(\vabth \rmvh,t), \label{Fh0}
    \\
      \Hho \left(\vvbth,t \right) &=& \left (I_{0,0} + J_{1,0} \right) / \vvbth , \label{Hho0D1V} \\
      \Gho \left(\vvbth,t \right) &=& \vvbth \left (\frac{I_{2,0} + J_{1,0}}{3} + I_{0,0} + J_{-1,0} \right)  ~. \label{Gho0D1V}
\end{eqnarray}
The symbols $\IiFho$ and $\JiFho$ denote the functionals of
$\Fho\left(\rmvhb,t \right)$, similar to
Shkarofsky et al.\cite{wang2024Relaxationmodel,Shkarofsky1967} , reads
  \begin{eqnarray}
      \IiFho \left(\vvbth,t \right) &=&
      \frac{1}{(\vvbth)^i} \int_0^{\vvbth} \left(\rmvhb \right)^{i+2} \Fho\left(\rmvhb,t \right) \rmd \rmvhb,  \quad i = L,L+2 , \label{IjFo} 
      \\
      \JiFho \left(\vvbth,t \right) &=& (\vvbth)^i \int_{\vvbth}^{\infty} \frac{\left(\rmvhb \right)^2}{\rmvhb^i} \Fho\left(\rmvhb,t \right) \rmd \rmvhb,  \quad i = L \pm 1  , \ \label{JjFo}
  \end{eqnarray}
where the normalized speed of species $b$, $\rmvhb = v_b / \vbth$. The definitions given by Eqs.\EQ{IjFo}-\EQo{JjFo} do not include the coefficient $4 \pi$. This coefficient arises from the application of spherical-polar coordinates in velocity space and is included in Eq.\EQ{collaba}. Similarly, the Jacobian $(\rmvh_b)^2$ in Eqs.\EQ{IjFo}-\EQo{JjFo} also originates from the utilization of spherical-polar coordinates.

\section{General relaxation model for homogeneous plasmas}
\label{General relaxation model for homogeneous plasmas}

By introducing a new set functions named $R$ (given in Appendix\APP{R function}) and $R$ integration (given in Appendix\APP{R integration}), and applying the FDIF hypothesis introduced in our previous paper\cite{wang2024Relaxationmodel}, a general relaxation model for homogeneous plasmas with spherically symmetric velocity space, including shell structures\cite{wang2024Relaxationmodel, Min2015}, will be presented based on the VFP spectrum equation\EQ{VFPl}.

\subsection{Transport equation}
\label{Transport equation}

In the spherical-polar coordinate system, the transport equations can be readily  derived\cite{wang2024Relaxationmodel} by multiplying $4 \pi m_a \rmv^{j+2} \rmd \rmv$ to both sides of the VFP spectrum equation\EQ{VFPl}, integrating over interval $\left [0, \infty \right)$, thereby resulting in the $(j,0)^{th}$-order kinetic moment evolution equation (KMEE)
\begin{eqnarray}
    \ddt \calMjo \left(t \right) &=& \rho_a \left(\vath \right)^{j} \calRhjo , \quad j \ge -2 , \label{dtMho}
\end{eqnarray}
where the mass density $\rho_a = m_a n_a$. That is the transport equation for case with spherically symmetric velocity space. In above equation, $\calMjo$ represents the $(j,0)^{th}$-order kinetic moment\cite{wang2024Relaxationmodel} given by
\begin{eqnarray}
    \calMjo \left(t \right) &=&  \rho_a \left(\vath \right)^{j} \calMhjo = 4 \pi \rho_a (\vath)^j \int_0^{\infty} \rmvh^{j+2} \fho \rmd \rmvh~. \label{Mjl0D1V}
\end{eqnarray}
Symbol $\calRhjo$ denotes the $(j,0)^{th}$-order normalized kinetic dissipative force
\begin{eqnarray}
    \calRhjo \left(t \right) &=& 4 \pi
    \int_0^{\infty} \rmvh^{j+2} \colhoa \rmd \rmvh, \label{Rjl0D1V}
\end{eqnarray}
where $j \ge -2$. Function $\calMhjo$ represents the $(j,0)^{th}$-order normalized kinetic moment. 

Especially when $j=0$ and $j=2$, the kinetic moments\EQ{Mjl0D1V} can be simplified as follows:
\begin{eqnarray}
    \calMoo \left(t \right) =  \rho_a , \label{rhoa}
    \quad
    \calM_{2,0} \left(t \right) = 2 K_a, \label{Ka}
\end{eqnarray}
where $K_a = (m_a / 2) \int_{\v} \v^2 f \rmd \v$, representing the total energy of species $a$. Thus, the transport equation\EQ{dtMho} represents the mass and energy conservation equations\cite{wang2024Relaxationmodel} respectively, and are stated as
\begin{eqnarray}
    {\frac{\partial}{\partial t}} \rho_a \left(t \right) &=& \rho_a {{\hat{\mathcal{R}}}{_{0,0}}}, \label{dtna}
    \\
    {\frac{\partial}{\partial t}}  K_a \left(t \right) &=& \frac{1}{2} \rho_a \left({{v}_{ath}} \right)^{2} \hat{\mathcal{R}}_{2,0} ~.\label{dtKa}
\end{eqnarray} 
The average velocity is zero, which is dependent on the amplitude of the distribution function when $l=1$. Therefore, the momentum conservation equation for scenarios where velocity space exhibits spherical symmetry will be
\begin{eqnarray}
    {\frac{\partial}{\partial t}} I_a \left(t \right) &=& \frac{1}{3} \rho_a {{v}_{ath}} {{\hat{\mathcal{R}}}{_{1,1}}} \ \equiv \ 0 ~. \label{dtIa}
\end{eqnarray}

Due to the symmetries\cite{Braginskii1965} during the Coulomb collision between two species, the normalized FPRS collision operator\EQ{collaba} ensures the conservation of mass, momentum, and energy theoretically
  \begin{eqnarray}
      n_a {{{\mathcal{\hat{R}}}{_{ab}}}{_{0,0}}} &=& n_b {{{\mathcal{\hat{R}}}{_{ba}}}{_{0,0}}} \ = \ 0, \label{Cnh0D1V}
      \\
      \frac{1}{3} \rho_a \vath {{{\mathcal{\hat{R}}}{_{ab}}}{_{1,1}}} &=& - \frac{1}{3} \rho_b \vbth {{{\mathcal{\hat{R}}}{_{ba}}}{_{1,1}}} \ = \ 0, \label{CIh0D1V}
      \\
      \frac{1}{2} \rho_a \vath^2 {\mathcal{\hat{R}}}{_{ab}}_{2,0} &=& - \frac{1}{2} \rho_b \vbth^2 {\mathcal{\hat{R}}}{_{ba}}_{2,0}~. \label{CKh0D1V}
  \end{eqnarray}
These are the conservation equations when the velocity space exhibits spherical symmetry.

By employing the relation $\rho_a (t) = m_a n_a$ and $K_a (t) = (3/2) n_a T_a$, expressions for number density and temperature can be derived in cases where velocity space exhibits spherical symmetry, reads
\begin{eqnarray}
    n_a (t) = {\calMoo}/{m_a},
    \quad
    T_a (t) = {\calM_{2,0}} / (3 n_a)  ~.
\end{eqnarray}
Subsequently, the thermal velocity can be determined according to the relation $\vath (t) = \sqrt{2 T_a / m_a}$, given by
\begin{eqnarray}
    \vath (t) &=& \sqrt{{2 \calM_{2,0}} / (3 \calMoo ) } ~.
\end{eqnarray}

As stated in the introduction and demonstrated in reference\cite{wang2024Relaxationmodel}, the aforementioned transport equation\EQ{dtMho} is characterized by inherent nonlinearity and a lack of closure. This is ascribed to two primary factors: I) The nonlinear interdependence among kinetic moments of different orders. II) The nonlinear dependence of the normalized kinetic dissipative forces on the kinetic moments. The closure strategies for these two aspects are provided in the following section.

\subsection{Closure relations}
\label{Closure relations}

Unlike the conventional approaches which usually simplify the nonlinear relations based on the near-equilibrium assumption\cite{Grad1949}, the set of transport equations has been truncated based on the FDIF hypothesis\cite{wang2024Relaxationmodel} in $(j)$ space. By employing the King function expansion\EQ{KFE} method, we establish closure relations based on the FDIF hypothesis for kinetic moments with CPEs (details in Sec.\SEC{Characteristic parameter equation}), as well as for the kinetic dissipative forces $\calRhjo$ with a kinetic dissipative force closure relation (details in Sec.\SEC{Kinetic dissipative force closure relation}). Furthermore, a kinetic moment-closed model is presented for homogeneous plasmas in the presence of spherical symmetry and shell structures in velocity space.

\subsubsection{King function expansion}
\label{King function expansion}

The spherical symmetry of the velocity space does not necessarily indicate that the system is in a thermodynamic equilibrium state. In this general scenario, Ref.\cite{wang2024Relaxationmodel} introduces a new function, which is referred to as zeroth-order King function as follows:
    \begin{eqnarray}
        \Ko \left(\rmvh;\iota, \sigma \right)  &=&  \frac{1}{\sqrt{2 \pi}} \frac{1}{\sigma^3 } \frac{\sigma^2}{2 \iota \rmvh } \exp{\left(-\frac{\rmvh^2 + \iota^2}{\sigma^2} \right)} \sinh{\left(\frac{2 \iota \rmvh }{\sigma^2} \right)} ,  \label{K0}
    \end{eqnarray}
where $\iota$ and $\sigma$ represent the expectation and deviation of King function. Under the FDIF hypothesis, by applying the King function expansion (KFE) technology\cite{wang2024Relaxationmodel}, the amplitude function $\fho$ can be approximated as
    \begin{eqnarray}
        \fho \left(\rmvh,t \right)  &=& \frac{\sqrt{2 \pi}}{\pi^{3/2}} \sum{_{r=1}^{\NKa}} \nhar \Ko \left(\rmvh;\uzhar,\vhathr \right)
        ,  \label{KFE}
    \end{eqnarray}
where $\NKa \in \bbN^+$. Here, 
$\nhar=\nar / n_a$, $\uzhar=\uzar / \vath$ and $\vhathr = \vathr / \vath$, representing the characteristic parameters of $r^{th}$ sub-distribution of $\fho$.
Similarly, the normalized amplitudes of the background distribution function of species $b$ can be approximated as
    \begin{eqnarray}
        \Fho \left(\rmvh_b,t \right)  &=& \frac{\sqrt{2 \pi}}{\pi^{3/2}} \sum{_{s=1}^{\NKb}} \nhbs \Ko \left(\rmvh;\uzhbs,\vhbths \right)
        ~.  \label{FhoFhos}
    \end{eqnarray}

  \begin{figure}[htbp]
	\begin{center}
		\includegraphics[width=0.65\linewidth]{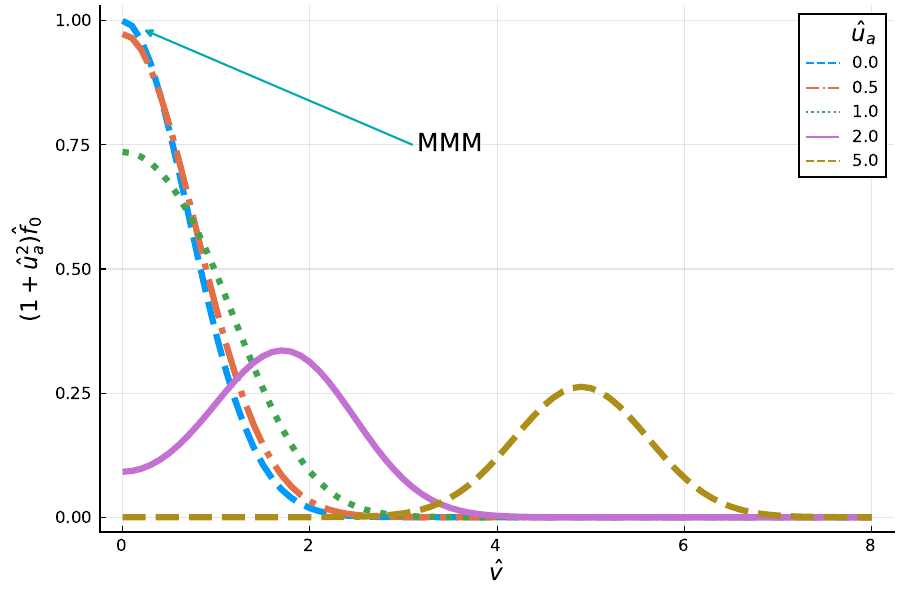}
	\end{center}
	\caption{Illustration of the amplitude functions multiplied by a factor $(1+\hat{u}_a^2)$ for $\NKa \equiv 1$ and various normalized average velocity $\hat{u}_a$ in KMM0.}
	\label{FigKMM0}
  \end{figure}
  
The FDIF hypothesis indicates that $\NKa$ is a finite-size number in the zeroth-order King mixture model (KMM0) represented by Eq.\EQ{KFE}. Let KMM0 denote the plasmas being in a quasi-equilibrium state\cite{wang2024higherorder}. Particularly, when $\sum_r(|{\hat{u}_{a_r}}|^2) \equiv 0$, KMM0 reduces to be the Maxwellian mixture model (MMM), indicating that the plasmas are in a shell-less quasi-equilibrium state\cite{wang2024Relaxationmodel}. The convergence of KMM0 can be established by using Wiener's Tauberian theorem\cite{Wiener1932,Korevaar2004} as referenced in Ref.\cite{wang2024Relaxationmodel}, and it has been demonstrated to have a moment convergence in Ref.\cite{wang2024Aconservative}.
When species $a$ is in states of quasi-equilibrium with $\NKa = 1$,  the amplitude functions described by KMM0 as functions of $\hat{v}$, along with various normalized average velocities $\hat{u}_a$, are depicted in Fig.\FIG{FigKMM0}. In order to examine the details when $\hat{u}_a > 1$, the amplitude functions are multiplied by a factor of $(1+\hat{u}_a^2)$.

\subsubsection{Characteristic parameter equation}
\label{Characteristic parameter equation}

Under FDIF hypothesis, by substituting KMM0\EQ{KFE} into the definition of normalized kinetic moments\EQ{Mjl0D1V} and simplifying the result, the characteristic parameter equations (CPEs) are obtained when velocity space exhibits spherical symmetry, namely,
      \begin{eqnarray}
            \calMhjo \left(t \right) &=& {C_M}_j^0 \sum{_{r=1}^{\NKa}} \nhar \left(\vhathr \right)^j \IFI
            \left [- \frac{j}{2}, \frac{3}{2}, -\left(\frac{\uzhar}{\vhathr} \right)^2 \right], \label{CPEs9Ms}
      \end{eqnarray}
where $j \ge -2$ and $\IFI (a,b,z)$ represents the Kummer confluent hypergeometric $\IFI$ function, and the coefficient
      \begin{eqnarray}
          {C_M}_j^0  &=&  \frac{2}{\sqrt{\pi}} \Gamma \left(\frac{j+3}{2} \right)~.   \label{CMj0} 
      \end{eqnarray}
Especially when $j$ is even, the aforementioned CPEs can be reformulated into a more concise and efficient form
      \begin{eqnarray}
            \calMhjo \left(t \right) = {C_M}_j^0 \sum_{r=1}^{\NKa} \nhar \left(\vhathr \right)^j 
            \left [1 + \sum_{\beta=1}^{j/2} C_{j,0}^\beta \left (\frac{\uzhar}{\vhathr} \right)^{2 \beta} \right], \label{CPEs9Ms0D1V}
      \end{eqnarray}
where $j \in \left\{(2j_p-2)|j_p \in \bbN^+ \right\}$. The coefficient 
      \begin{eqnarray}
          C_{j,0}^\beta  &=&  2^\beta {C_\beta^{j/2}} / {(2 \beta + 1)!!},
      \end{eqnarray}
and $C_\beta^{j/2}$ represents the binomial coefficient.
 
The CPEs\EQ{CPEs9Ms} with various values of $j$ constitute a set of nonlinear algebraic equations. 
Generally, these equations typically involve a total of $3 \NKa$ unidentified parameters, which can be determined by solving the well-posed CPEs with $3 \NKa$ known normalized kinetic moments $\calMhjo$. This approach is delineated in Ref.\cite{wang2024Relaxationmodel,wang2024Aconservative}.
From the perspective of spectral analysis (where the King function serves as a non-orthogonal basis), distinct King functions represent different spectral components, while the characteristic parameters are quantities in the spectral space $(j)$ rather than the physical quantities. This study is focus on offering the closure relations for transport equations\EQ{dtMho}, providing kinetic moment-closed model for case with spherically symmetric velocity space.

\subsubsection{Kinetic dissipative force closure relation}
\label{Kinetic dissipative force closure relation}

To enclose the kinetic dissipative forces arising from the collision operator\EQ{collaba}, one can rephrase Eq.\EQ{Rjl0D1V} as
\begin{eqnarray}
    \calRhjo \left(t \right) &=& \sum_b \nbvth \Gab \calRhabjo  , \label{Rhj0a0D1V}
\end{eqnarray}
where
\begin{eqnarray}
    \calRhabjo \left(t \right)  =  4 \pi 
    \int_0^{\infty} \rmvh^{j+2} \colhoab \rmd \rmvh , \quad j \ge -2 ~. \label{Rhabjl0D1V}
\end{eqnarray}
Let Eq.\EQ{Rhj0a0D1V} represent the kinetic dissipative force closure relation.

The analytical expression of the kinetic dissipative force closure relations can be derived by substituting KMM0\EQ{KFE} into FPRS collision operator as described by Eq.\EQ{collaba}, and then applying Eq.\EQ{Rhabjl0D1V}. With the introduction of a new set functions, including the $R$ function (details in Appendix\SEC{R function}), the normalized kinetic dissipative forces between species $a$ and  species $b$ can be formulated as
\begin{eqnarray}
    \calRhabjo \left(t \right) &=& \frac{\sqrt{2 \pi}}{{\pi^{3/2}}} \sum_{s=1}^{\NKb} \nhbs  \sum_{r=1}^{\NKa} \nhar \calRhabrsjo,  \label{Rhabjl0D1VKing}
\end{eqnarray}
where
  \begin{eqnarray}
     \calRhabrsjo \left(t \right)   &=&  
     \IhSabrsjo +  \IhCabrsjo +  \IhDabrsjo, \quad j \ge -2   ~ . \label{Rhabrsj000nuTs9Ms0D1V}
  \end{eqnarray}
The terms on the right-hand side of above equation correspond to the $R$ integration, as defined in Appendix\SEC{R integration}.


Especially, the utilization of the outcomes of the $R$ function when the velocity space exhibits spherical symmetry without shell structure\cite{wang2024Relaxationmodel}, as represented by Eqs.\EQ{n0A0u0}-\EQo{n1u0} ( wherein $\sum_r(|\uzhar|^2)  \equiv 0$  and $\sum_s(|\uzhbs|^2)  \equiv 0$), Eq.\EQ{Rhabjl0D1VKing} is reduced to the form presented in our previous work\cite{wang2024Relaxationmodel}, which reads
\begin{eqnarray}
  \begin{aligned}
    \calRhabjo \left(t \right) \ =& \ \frac{2}{\pi^2} \rmGamma \left(\frac{j+3}{2} \right) \sum_{s=1}^{\NKb} \nhbs  \sum_{r=1}^{\NKa} \nhar \frac{1}{\vhathr^7}  \left [ m_M \vhathr \left(\frac{1 + \vabth^2}{\vhathr^2} \right)^{-(j+3)/2} - 
      \right. \\& \left.
    \left(\vhathr \right)^{j+3} \frac{\vhbths^2-m_M \vabth^2 \vhathr^2}{\vabth^4 \vhbths} \IIFI \left(\frac{1}{2},\frac{j+3}{2},\frac{3}{2},-\frac{\vabth^2 \vhathr^2}{\vhbths^2}\right) +  \ 
    \right. \\& \left.
    \left (m_M - 1 - \frac{\vhbths}{\vabth^4} \right) \frac{\vhathr^2}{\vabth^2 \vhbths} \left(\frac{1}{\vhathr^2} + \frac{\vabth^2}{\vhbths^2}\right)^{-(j+3)/2}
      \right],  \quad j \ge -2
    ~. \label{Rhabj000D1V}
  \end{aligned}
\end{eqnarray}
Here, $\IIFI(a,b,c,z)$ denotes the Gauss hypergeometric $\IIFI$ function\cite{Arfken1971} with variable $z$.
  
\subsection{Kinetic moment-closed model based on KMM0}
\label{Kinetic moment-closed model based on KMM0}

Under FDIF hypothesis and by employing the KFE\EQ{KFE} method for speed coordinate, the evolution of the plasmas system can be described by the transport equation\EQ{dtMho}, incorporating the analytical expression\EQ{Rhabjl0D1VKing} of the normalized kinetic dissipative forces between species $a$ and  species $b$.
The conservation equations represented by Eqs.\EQ{dtna}-\EQo{dtIa}, the characteristic parameter equations\EQ{CPEs9Ms} and the kinetic dissipative force closure relations\EQ{Rhj0a0D1V} serves as the constraint equations of the transport equation\EQ{dtMho}.
The combination of these equations constitutes a set of nonlinear equations, which will be referred to as \textbf{kinetic moment-closed model (KMCM)} for the scenarios in which velocity space exhibits spherical symmetry. The advantages of KFE, and subsequently, the benefits of KMCM are outlined in Ref.\cite{wang2024Relaxationmodel, wang2024higherorder}.

In general, this model provides a nonlinear nonequilibrium description of the kinetic moments by using characteristic parameters as intermediate parameters, serving as a general relaxation model for homogeneous plasmas with spherically symmetric velocity space. 
For numerically solving this nonlinear model, the transport equation\EQ{dtMho} can be solved\cite{wang2024Aconservative} by utilizing a Runge-Kutta solver, such as the trapezoidal scheme\cite{Rackauckas2017}. The characteristic parameter equations\EQ{CPEs9Ms} can be addressed\cite{wang2024Aconservative} through a least squares method\cite{Fong2011} (LSM). As mentioned in our previous paper\cite{wang2024Relaxationmodel} and demonstrated in a 0D-2V situation\cite{wang2024Aconservative}, a general optimization algorithm, including self-adaptive $\NKa$ and collection of $j$, must be adopted for the solution of this nonlinear system. This will be addressed in our future research.


\section{Temperature relaxation model}
\label{Temperature relaxation model}

One potential application of the aforesaid general relaxation model is to establish a temperature relaxation model for homogeneous plasmas with general spherically symmetric velocity space.
When $j=0$, the mass conservation equation can be determined by substituting relation represented by Eq.\EQ{rhoa} into the transport equation\EQ{dtMho}, reads
\begin{eqnarray}
    \ddt n_a \left(t \right)  &=&  0 ~.\label{dtna0D1V}
\end{eqnarray}
The above equation utilizes the fact that $\calRhaboo \equiv 0$ for all elastic collisions given by Eq.\EQ{Cnh0D1V}.

When $j=2$, the energy conservation equation can be derived by substituting the relation $\calM_{2,0} \left(t \right) =  2 K_a$\EQ{Ka} into the transport equation\EQ{dtMho}, as follows: 
\begin{eqnarray}
    \ddt K_a \left(t \right)  &=&  \frac{3}{4}  \rho_a \vath ^{2} \sum_b \nbvth \Gab \calRhab_{2,0} ~.\label{dtKa0D1V}
\end{eqnarray}
Neglecting relativistic effects, the application of the relation $K_a = \frac{3}{2} n_a T_a$, and mass conservation\EQ{dtna0D1V} results in a temperature relaxation equation,
\begin{eqnarray}
    \ddt T_a \left(t \right)  &=& - \nuTa T_a~. \label{dtTa0D1V}
\end{eqnarray}
The general characteristic frequency of temperature relaxation,
\begin{eqnarray}
    \nuTa \left(t \right)  &=& - \frac{2}{3} \sum_b \nbvth \Gab \calRhab_{2,0} ~.\label{nuTtab0D1V}
\end{eqnarray}
Eq.\EQ{dtTa0D1V} serves as a general temperature relaxation model for scenario where velocity space exhibits spherical symmetry. Generally, $\nuTa$ shows a nonlinear dependent on $T_a$ and $T_b$. The solution to the above equation\EQ{dtTa0D1V} will take the following form only when $\nuTa$ has weak dependence on $T_a$,
\begin{eqnarray}
    T_a \left(t \right) &=& T_{a_0} \e^{- \nuTa t}, \label{TatD1V}
\end{eqnarray}
where $T_{a_0} = T_a (t =0)$ which will be determined by the initial condition.

Our previous investigation\cite{wang2024Relaxationmodel} has revealed that when the near-equilibrium condition is fulfilled during two-species collisions, the temperature relaxation equation\EQ{dtTa0D1V} can be explicitly reformulated as the traditional zeroth-order Braginskii heat transfer model\cite{wang2024Relaxationmodel,Braginskii1958}, assuming that both species are in close proximity to their respective thermodynamic equilibrium states. This zeroth-order model ignores the deviation of distribution function from the Maxwellian one and can be expressed as
  \begin{eqnarray}
      \ddt T_a \left(t \right) &=& - \nuTab \left (T_a - T_b \right),  \label{dtTa9Ms0D1V}
  \end{eqnarray}
where $\nuTab$ represents the characteristic frequency of temperature relaxation as given by Huba\cite{Huba2011}, and also derived from the shell-less KMCM\cite{wang2024Relaxationmodel}. This characteristic frequency is quoted as follows:
\begin{eqnarray}
    {\nu_T^{ab}} &=& \frac{8\sqrt{2\pi}}{3} \left(\frac{q_e^2}{4\pi \varepsilon_0} \right)^2 \frac{\sqrt{m_a m_b} \left(Z_a Z_b \right)^2 n_b}{\left(m_a T_b + m_b T_a \right)^{3/2}} {\ln{ \Lambda_{ab}}}  ~. \label{nuTab}
\end{eqnarray}

However, the function $\calRhab_{2,0}$\EQ{Rhabjl0D1VKing} is generally a nonlinear function of $m_M$, $\vabth$ and characteristic parameters of species $a$ and species $b$. By comparing Eq.\EQ{dtTa9Ms0D1V} and Eq.\EQ{dtTa0D1V}, the characteristic frequency of temperature relaxation in the presence of spherical symmetry in velocity space can generally be expressed as
\begin{eqnarray}
    \nuTab &=& \frac{T_a}{T_a - T_b} \nuTa
    ~. \label{nuTab00D1V}
\end{eqnarray}
The characteristic frequency of temperature relaxation described by Huba\cite{Huba2011} is clearly a specific instance of the one described by Eq.\EQ{nuTab00D1V}.
However, when $T_a$ is very close to $T_b$ for a quasi-equilibrium state plasma, the representation of $\nuTab$ by Eq.\EQ{nuTab00D1V} may numerically tend towards infinity. In this situation,  Eq.\EQ{dtTa9Ms0D1V} should be replaced by the original temperature relaxation model represented by Eq.\EQ{dtTa0D1V}.

It is more evident to observe the distinctions in the physics pictures between the general temperature relaxation model\EQ{dtTa0D1V} and the zeroth-order Braginskii heat transfer model\EQ{dtTa9Ms0D1V}, particularly in the simple scenario where the velocity space exhibits spherical symmetry without shell structure\cite{wang2024Relaxationmodel}.
Under the traditional near-equilibrium assumption, if $T_a$ is very close to $T_b$, the plasmas system tends towards the thermodynamic equilibrium state in the zeroth-order Braginskii heat transfer model.
However, under FDIF hypothesis in the presented general temperature relaxation model, the plasmas system may still be in a shell-less quasi-equilibrium state\cite{wang2024Relaxationmodel}.

\section{Conclusion}
\label{Conclusion}

In this paper, we propose a general relaxation model, namely a kinetic moment-closed model (KMCM) for homogeneous plasmas with general spherically symmetric velocity space. 
This model consists of the transport equation and normalized kinetic dissipative forces between two species, along with incorporated constraint equations including the conservation equations, characteristic parameter equations (CPEs) and kinetic dissipative force closure relation. 
By employing the KMM0 which is based on the finitely distinguishable independent features (FDIF) hypothesis, and introducing a set of special functions named as $R$ function and $R$ integration, the closure relations of kinetic moments and kinetic dissipative forces can be presented in a closed form. 
Furthermore, in cases with shell-less velocity space, this closure relation for kinetic dissipative forces reduces to the simple version presented in our precious work, which is expressed in terms of Gauss hypergeometric $\IIFI$ function. 

The presented general relaxation model accurately captures both the near-equilibrium state and the far-from-equilibrium states of homogeneous plasmas with spherically symmetric velocity space. 
These findings offer a new semi-analytical, semi-numerical approach for solving the VFP equation. The general temperature relaxation model is also achieved, and the general characteristic frequency of temperature relaxation in the presence of spherical symmetry in velocity space can be explicitly determined.
Furthermore, this general relaxation model can serve as valuable benchmarks for nonlinear statistical physics applications such as fusion plasmas and solar plasmas. 

In our upcoming research, we aim to address the general optimization algorithm for solving this nonlinear system, and extend these findings to encompass scenarios involving axisymmetric velocity space. Subsequently, this nonlinear model can be employed to investigate the classical physics phenomena such as Landau damping and Bernstein instability driven by the shell-type ion velocity distribution.

\section{Acknowledgments}
\label{Acknowledgments}


We express our gratitude to Jianyuan Xiao, Zhihui Zou, Pengfei Zhang , Heyi Li, Yifeng Zheng, Xianhao Rao and Wenlu Zhang for their valuable contributions to the discussions. We also extend our special thanks to Ge Zhuang, Jinlin Xie and Guanghui Zhu for their supports.
This work is supported by the NSFC (No.120051410).

\appendix    

\begin{appendices}

\section{$R$ function}
\label{R function}

 Similar to the hypergeometric function\cite{Arfken1971}, a set of new functions is defined, which are named as the $R$ function. The integral form can be expressed as
 \begin{eqnarray}
     _{j}\rmR_0 &  \left (\gamma,\iota_1,\sigma_1,\iota_2,\sigma_2;n, \alpha \right)  \ = \ \quad \quad \quad \quad \quad \quad  \quad \quad \quad \quad \quad \quad \quad  \quad \quad  \quad \quad \quad \quad  \quad \quad \quad \quad \quad \nonumber
     \\  &
     \left \{
  \begin{aligned}
      \int_0^\infty \rmvh^{j+2} \Ko \left(\gamma \rmvh;\iota_2,\sigma_2 \right) \Ko \left(\rmvh;\iota_1,\sigma_1 \right) \rmd \rmvh, \quad n=0,\alpha = 0, \quad \quad \quad \quad \quad \quad \quad \label{RKL}
      \\
     \int_0^\infty C_{0,s}^{n, \alpha} \rmvh^{j+2-n} \exp{\left[- \left(\frac{\gamma \rmvh \pm \iota_2}{\sigma_2} \right)^{2} \right]} \Ko \left(\rmvh;\iota_1,\sigma_1 \right) \rmd \rmvh, \quad n=0,2,\alpha = \pm 1, \label{RexpIL}
      \\ 
     \frac{1}{2 \gamma} \int_0^\infty C_{0,s}^{n, \alpha} \rmvh^{j+1} \exp{\left[- \left(\frac{\gamma \rmvh \pm \iota_2}{\sigma_2} \right)^{2} \right]} \Ko \left(\rmvh;\iota_1,\sigma_1 \right) \rmd \rmvh, \quad n=\pm 1,\alpha = \pm 1   \label{RexpJpL0}
  \end{aligned}  
  \label{RexpJ}
      \right.
 \end{eqnarray}
 and
 \begin{eqnarray}
     _{j} \rmR_0   & \left (\gamma,\iota_1,\sigma_1,\iota_2,\sigma_2;n \right) \ = \ \quad \quad \quad \quad \quad \quad  \quad \quad \quad \quad \quad \quad \quad  \quad \quad \quad \quad \quad \quad \quad \quad \quad \quad \quad\nonumber
     \\  &
     \left \{
  \begin{aligned}
     \int_0^\infty \rmvh^{j+2-n} \left[\erf \left(\frac{\gamma \rmvh + \iota_2}{\sigma_2} \right) + \erf \left(\frac{\gamma \rmvh - \iota_2}{\sigma_2} \right) \right] \Ko \left(\rmvh;\iota_1,\sigma_1 \right) \rmd \rmvh, \quad n = 0, 2, \label{RerfIL2}
     \\
     \int_0^\infty \rmvh^{j+2-1} \left[\erf \left(\frac{\gamma \rmvh + \iota_2}{\sigma_2} \right) - \erf \left(\frac{\gamma \rmvh - \iota_2}{\sigma_2} \right) \right] \Ko \left(\rmvh;\iota_1,\sigma_1 \right) \rmd \rmvh , \quad n = \pm 1, \label{RerfJLp1}
  \end{aligned}  
     \right. \quad \quad \label{Rerf}
 \end{eqnarray}
 where $j \ge - 2$ {and the coefficients $C_{0,s}^{n,+1}$, $C_{0,s}^{n,-1}$ in Eq.\EQ{RexpJ} will be}
  \begin{eqnarray}
      C_{0,s}^{n,\pm 1} &=& \left[1, \uzhbs, (\uzhbs)^2, \cdots, (\uzhbs)^{\frakn} \right] \Cn_{0,s}^{n,\pm 1} \left[1, \vvbth, (\vvbth)^2, \cdots, (\vvbth)^{\frakm} \right]^{\rmT} ~. \label{CjI}
  \end{eqnarray}
{Functions $\Cn_{0,s}^{n,\pm 1}$ are both matrices of size $\frakn \times \frakm$, which can be expressed as}
  \begin{eqnarray}
      \Cn_{0,s}^{n,\pm 1} &=& \delta_n^{1} 
      \begin{bmatrix}
      0 
      \end{bmatrix}
      +
       \delta_n^{-1}
      \begin{bmatrix}
      0  & \mp 1 \\
      1  & 0  \\
      \end{bmatrix} 
      + 
      \delta_n^0 
      \begin{bmatrix}
      \pm 1 
      \end{bmatrix}
      +
       \delta_n^{2}
      \begin{bmatrix}
      \pm \left(\vhbths \right)^2 & 0    &  \pm 1  \\
      0   & -1   &  0  \\
      \pm 1   & 0    &  0  \\
      \end{bmatrix} ~.
       \label{CjpIJLF0L0}
  \end{eqnarray}

\subsection{Special cases of $R$ function for KMM0}

When $n=\alpha=0$, the $R$ function can be expressed in terms of Kummer confluent hypergeometric $\IFI$ function\cite{Arfken1971}, given by
  \begin{eqnarray}
  \begin{aligned}
     _{j}\rmR_0 \left (\gamma,\iota_1,\sigma_1,\iota_2,\sigma_2;0,0 \right)  \ = & \ \frac{1}{32 \pi \gamma \sigma_1 \sigma_2 \iota_1 \iota_2} \left(\sigma_{12} \right)^{\frac{-j-1}{2}} \Gamma\left(\frac{j+1}{2} \right) 
     \exp{\left(- \frac{\iota_1^2}{\sigma_1^2}- \frac{\iota_2^2}{\sigma_2^2} \right)}
     \\&
    \times \left \{- \IFI \left[\frac{j+1}{2}, \frac{1}{2}, \left(\xi_{12}^- \right)^2 \right] 
     + 
     \IFI \left[\frac{j+1}{2}, \frac{1}{2}, \left(\xi_{12}^+ \right)^2 \right]
     \right \}
     ,
     \label{RKL1F1n0A0}
 \end{aligned}
 \end{eqnarray}
 where
  \begin{eqnarray}
      \sigma_{12} &=& \frac{1}{\sigma_1^2} + \frac{\gamma^2}{\sigma_2^2} , \label{sigma12}
      \quad
      \xi_{12}^{\pm} \ = \ \frac{\sigma_2^2 \iota_1 \pm \gamma \sigma_1^2 \iota_2 }{\sigma_1 \sigma_2 \sqrt{\sigma_2^2 + \gamma^2 \sigma_1^2}} ~. \label{xi12p}
  \end{eqnarray}
  When $n=0$ and $\alpha=\pm 1$, the $R$ function will be
  \begin{eqnarray}
  \begin{aligned}
     _{j}\rmR_0 & \left (\gamma,\iota_1,\sigma_1,\iota_2,\sigma_2;0, \alpha \right)  \ = \ \frac{\alpha \sigma_2}{8 \sqrt{2 \pi} \iota_1  \sqrt{\sigma_2^2 + \gamma^2 \sigma_1^2}} \left(\sigma_{12} \right)^{\frac{-j-1}{2}} \exp{\left(- \frac{\iota_1^2}{\sigma_1^2}- \frac{\iota_2^2}{\sigma_2^2} \right)}  
     \\&
     \times \left(
     \Gamma \left(\frac{j}{2}+1 \right) 
     \left \{ \IFI \left[\frac{j}{2}+1, \frac{1}{2}, \left(\xi_{12}^- \right)^2 \right] 
     + 
     \IFI \left[\frac{j}{2}+1, \frac{1}{2}, \left(\xi_{12}^+ \right)^2 \right]
     \right \}
     \right. \\&  \left.
     + 2 \sigma_1 \sigma_2  \Gamma{\left(\frac{j+3}{2} \right)}
     \left \{\xi_{12}^- \IFI \left[\frac{j+3}{2}, \frac{3}{2}, \left(\xi_{12}^- \right)^2 \right] 
     + 
     \xi_{12}^+ \IFI \left[\frac{j+3}{2}, \frac{3}{2}, \left(\xi_{12}^+ \right)^2 \right]
     \right \}
      \right)
     ~.
     \label{RKL1F1n0Ap}
 \end{aligned}
 \end{eqnarray}
 Similarly, the other $R$ function, $_{j}\rmR_0  \left (\gamma,\iota_1,\sigma_1,\iota_2,\sigma_2;n, \pm 1 \right)$ can also be expressed in terms of the Kummer confluent hypergeometric $\IFI$ function when $n=2$ or $n=\pm 1$.

\subsection{Special cases of $R$ function for MMM}

 When $\iota_1 = \iota_2 \equiv 0$ for $\forall (n, \alpha)$, the function $_{j}\rmR_0  \left (\gamma,\iota_1,\sigma_1,\iota_2,\sigma_2;n, \alpha \right)$ can be expressed in terms of Gamma function\cite{Arfken1971}, reads 
  \begin{eqnarray}
      _{j}\rmR_0 \left (\gamma,\iota_1,\sigma_1,\iota_2,\sigma_2;0,0 \right) &=& \frac{\left (\sigma_1 \sigma_2\right)^3}{4 \pi}  \left (\sigma_{12} \right)^{-\frac{j+3}{2}} \Gamma{\left (\frac{j+3}{2} \right)}, \label{n0A0u0}
      \\ 
      _{j}\rmR_0 \left (\gamma,\iota_1,\sigma_1,\iota_2,\sigma_2;0,\pm 1 \right) &=& \frac{\pm 1}{2 \sqrt{2\pi} \sigma_1^3} \left (\sigma_{12} \right)^{-\frac{j+3}{2}} \Gamma{\left (\frac{j+3}{2} \right)}, \label{n0Au0}
      \\ 
      _{j}\rmR_0 \left (\gamma,\iota_1,\sigma_1,\iota_2,\sigma_2;2,\pm 1 \right) &=& \pm 1 \frac{2 \sigma_2^2 + (j+3) \gamma^2 \sigma_1^2}{4 \sqrt{2\pi} \sigma_1^3} \left (\sigma_{12} \right)^{-\frac{j+3}{2}} \Gamma{\left (\frac{j+1}{2} \right)}, \label{n2Au0}
      \\ 
      _{j}\rmR_0 \left (\gamma,\iota_1,\sigma_1,\iota_2,\sigma_2;1,\pm 1 \right) &=& \frac{\pm \gamma}{2 \sqrt{2\pi} \sigma_1^3} \left (\sigma_{12} \right)^{-\frac{j+5}{2}} \Gamma{\left (\frac{j+5}{2} \right)}, \label{npAu0}
      \\ 
      _{j}\rmR_0 \left (\gamma,\iota_1,\sigma_1,\iota_2,\sigma_2;-1,\pm 1 \right) & \equiv & 0 ~. \label{nnAu0}
  \end{eqnarray}
 Function $_{j} \rmR_0 \left (\gamma,\iota_1,\sigma_1,\iota_2,\sigma_2;n \right)$ for $\forall n$ can also be expressed in terms of Gauss hypergeometric $\IIFI$ function\cite{Arfken1971}, as follows:
  \begin{eqnarray}
      _{j}\rmR_0 \left (\gamma,\iota_1,\sigma_1,\iota_2,\sigma_2;0 \right) &=& \frac{\sqrt{2}}{2 \pi} \frac{\left (\sigma_1 \right)^{j+1}}{\sigma_2} \Gamma{\left (\frac{j}{2}+2 \right)} \IIFI \left[\frac{1}{2}, \frac{j}{2}+2,\frac{3}{2}, - \left (\sigma_r \right)^2 \right], \label{n0u0}
      \\ 
      _{j}\rmR_0 \left (\gamma,\iota_1,\sigma_1,\iota_2,\sigma_2;2 \right) &=& \frac{\sqrt{2}}{2 \pi} \frac{\left (\sigma_1 \right)^{j-1}}{\sigma_2} \Gamma{\left (\frac{j}{2}+1 \right)} \IIFI \left[\frac{1}{2}, \frac{j}{2}+1,\frac{3}{2}, - \left (\sigma_r \right)^2 \right], \label{n2u0}
      \\ 
      _{j}\rmR_0 \left (\gamma,\iota_1,\sigma_1,\iota_2,\sigma_2;\pm 1 \right) &=& \frac{\sqrt{2}}{2 \pi} \frac{\left (\sigma_1 \right)^{j}}{\sigma_2} \Gamma{\left (\frac{j+3}{2} \right)} \IIFI \left[\frac{1}{2}, \frac{j+3}{2},\frac{3}{2}, - \left (\sigma_r \right)^2 \right], \label{n1u0}
  \end{eqnarray}
 where $\sigma_r = \gamma \sigma_1 / \sigma_2$.

\section{$R$ integration}
\label{R integration}

 With the definition of $R$ functions given in Sec.\SEC{R function}, the $R$ integration in Eq.\EQ{Rhabrsj000nuTs9Ms0D1V} can be expressed as
  \begin{eqnarray}
      \IhSabrsjo \left(t \right) &=& \ 4 \pi \frac{\sqrt{2 \pi}}{\pi^{3/2}} m_M \times \ _{j} \rmR_0 \left (\vabth,\uzhar,\vhathr,\uzhbs,\vhbths; 0,0 \right),  \quad \label{RintSabR0L5}
  \end{eqnarray}
  \begin{eqnarray}
  \begin{aligned}
      \IhCabrsjo \left(t \right) \ = \ & m_M c_{0}^{erf} \times \ 
      \left. 
       _{j} \rmR_0 \left (\vabth,\uzhar,\vhathr,\uzhbs,\vhbths;0 \right) -
      \right. \\ &  \left.
      \frac{1}{3} c_{2}^{erf} \times \  _{j} \rmR_0 \left (\vabth,\uzhar,\vhathr,\uzhbs,\vhbths;2 \right)  +
      \right. \\ &  \left.
      \frac{2}{3} c_{1}^{erf} \times \ 
      _{j} \rmR_0 \left (\vabth,\uzhar,\vhathr,\uzhbs,\vhbths;+ 1 \right) +
      \right. \\ &  \left.
       m_M c_{I}^{exp} \times \ _{j} \rmR_0 \left (\vabth,\uzhar,\vhathr,\uzhbs,\vhbths;0,+1 \right) -
      \right. \\ &  \left.
      \frac{1}{3} c_{I}^{exp} \times \ _{j} \rmR_0 \left (\vabth,\uzhar,\vhathr,\uzhbs,\vhbths;2,+1 \right)  +
      \right. \\ &  \left.
      \frac{2}{3} c_{J}^{exp} \times \ _{j} \rmR_0 \left (\vabth,\uzhar,\vhathr,\uzhbs,\vhbths;1,+1 \right) +
      \right. \\ &  \left.
       m_M c_{I}^{exp} \times \ _{j} \rmR_0 \left (\vabth,\uzhar,\vhathr,\uzhbs,\vhbths;0,-1 \right) -
      \right. \\ &  \left.
      \frac{1}{3} c_{I}^{exp} \times \ _{j} \rmR_0 \left (\vabth,\uzhar,\vhathr,\uzhbs,\vhbths;2,-1 \right)  +
      \right. \\ &  \left.
      \frac{2}{3} c_{J}^{exp} \times \ _{j} \rmR_0 \left (\vabth,\uzhar,\vhathr,\uzhbs,\vhbths;1,-1 \right) 
       \right. 
  \end{aligned}  \label{RintCabRl05}
  \end{eqnarray}
and
  \begin{eqnarray}
  \begin{aligned}
      \IhDabrsjo \left(t \right) \ = \ & \frac{1}{3} \times 
      \left [c_{2}^{erf} \times \ 
      _{j} \rmR_0 \left (\vabth,\uzhar,\vhathr,\uzhbs,\vhbths;2 \right)  +
      \right. \\ &  \left. \quad
      2 c_{1}^{erf} \times \
      _{j} \rmR_0 \left (\vabth,\uzhar,\vhathr,\uzhbs,\vhbths;+ 1 \right) +
      \right. \\ &  \left.
      \quad c_{I}^{exp} \times \ _{j} \rmR_0 \left (\vabth,\uzhar,\vhathr,\uzhbs,\vhbths;2,+1 \right)  +
      \right. \\ &  \left. \quad
      2 c_{J}^{exp} \times \ _{j} \rmR_0 \left (\vabth,\uzhar,\vhathr,\uzhbs,\vhbths;1,+1 \right) +
      \right. \\ &  \left.
      \quad c_{I}^{exp} \times \
      _{j} \rmR_0 \left (\vabth,\uzhar,\vhathr,\uzhbs,\vhbths;2,-1 \right)  +
      \right. \\ &  \left. \quad
      2 c_{J}^{exp} \times \
      _{j} \rmR_0 \left (\vabth,\uzhar,\vhathr,\uzhbs,\vhbths;1,-1 \right) 
       \right],
  \end{aligned}  \label{RintDabRl05}
  \end{eqnarray}
where $j \ge -2$ and $\vabth = \vath / \vbth$. The coefficients in above equations are 
  \begin{eqnarray}
      c_{0}^{erf} &=& 4 \pi \frac{\sqrt{\pi}}{2^3} ,
      \quad
      c_{I}^{exp} \ = \ 4 \pi \frac{\vhbths \uzhbs}{2} ,
      \quad
      c_{J}^{exp} \ = \ 4 \pi \frac{\vhbths}{4 \uzhbs} 
  \end{eqnarray}
and
  \begin{eqnarray}
      c_{2}^{erf} &=& 4 \pi \frac{\sqrt{\pi}}{2^3}  \left(\vhbths \right)^{2}  \left[\frac{3}{2} + \left(\frac{\uzhbs}{\vhbths} \right)^{2} \right] \left(\frac{\vabth}{\vhbths}\right)^{-2},
      \\
      c_{1}^{erf} &=& 4 \pi \frac{\sqrt{\pi}}{2^3} \left[1 +\frac{1}{2} \left(\frac{\sigma_2}{\uzhbs} \right)^{2} \right] \left(\frac{\vabth}{\uzhbs} \right)^{-1} ~.
  \end{eqnarray}
When species $b$ is identical to species $a$, the  mass ratio and the thermal velocity ratio are both equal to one, that is, $m_M = \vabth \equiv 1$. Subsequently, the aforementioned functions can be reduced to the $R$ function for self-collision processes.
  

\end{appendices}

\
\
\ 
\
\
\ 
\

\end{spacing}
 
\begin{spacing}{0.5}  
 
\
\


\end{spacing}

\bibliographystyle{iopart-num.bst}
\bibliography{Plasma}

\end{document}